# Nonlocal spin transport in single walled carbon nanotube networks

Hyunsoo Yang,[1,*] Mikhail E. Itkis,[2] Rai Moriya,[3] Charles Rettner,[3] Jae-Seung Jeong,[4] Daniel Pickard,[1] Robert C. Haddon,[2] and Stuart S. P. Parkin[3]

[1] Department of Electrical and Computer Engineering, National University of Singapore, 4 Engineering Drive 3, 117576 Singapore

[2] Center for Nanoscale Science and Engineering, Departments of Physics and Astronomy, Chemistry and Chemical and Environmental Engineering, University of California, Riverside, California 92521

[3] IBM Almaden Research Center, 650 Harry Road, San Jose, California 95120

[4] School of Physics, Korea Institute for Advanced Study, Seoul 130-722, Korea

## Abstract

Spin transport in carbon-based materials has stimulated much interest due to their ballistic conductance and a long phase coherence length. While much research has been conducted on individual carbon nanotubes, current growth and placement techniques are incompatible with large-scale fabrication. Here we report nonlocal spin injection and detection in single wall carbon nanotube networks. We observe spin transport over a distance of 1 μm, and extract a spin diffusion length of 1.6 - 2.4 μm with an injected spin polarization from CoFe into nanotube network of 18 - 41%. Our observations demonstrate that spin transport is possible in carbon nanotube networks due to the formation of natural tunnel barriers between nanotubes and metallic contacts.

Spin based devices benefit greatly from long spin coherence times and distances. Among the spin transport media studied to date, semiconductors have the advantage of longer spin relaxation times compared to metals, but with smaller carrier velocities. In principle, carbon nanotubes (CNTs) are a promising alternative media for the transport of spin information, since they have weak hyperfine interactions, long spin relaxation times, and large spin travel velocities. Local and nonlocal magnetoresistances (MR) have been reported in single carbon nanotube nanowires.[1-11] However, the self assembly style growth of carbon nanotubes is not well suited to top-down device fabrication for mass production, yielding only a few working devices per wafer. Here, we show that single walled carbon nanotube networks, where the nanotubes are in the form of densely packed thin films, are appealing candidates for spintronic applications. We show that the resistance of ferromagnetically contacted nanotube network switches hysteretically as a function of applied magnetic field in a nonlocal geometry, which is due neither to anisotropic magnetoresistance nor to the anomalous Hall effect in the ferromagnetic leads.

For this work we utilized highly purified electric arc discharge produced single-walled carbon nanotubes (SWNTs) of average diameter 1.55±0.1 nm.[12, 13] SWNT thin films were prepared by vacuum filtration of a SWNT dispersion using an alumina filtration membrane disk (47 mm dia, 0.02 μm pore size, 34 mm working dia.) following procedures in the literature.[14, 15] The thickness of the SWNT films was controlled by the amount of the SWNT material utilized for the vacuum filtration, taking into account the working area of the filtration membrane; the final thickness calibration was obtained by use of a Dektak profilometer on thicker (> 200 nm) SWNT films. All the data in this manuscript were taken on films prepared from a dispersion of 0.060 mg of SWNT material in 100 mL of solvent with a resulting thickness of the SWNT film of 50 nm corresponding to a SWNT film density of 1.2 g/cm$^3$.[16] The SWNT film was separated from the alumina membrane by dipping into a 0.05N NaOH solution followed by the transfer of the free floating SWNT film in DI water. The SWNT film was then placed at the designated position on the patterned Si wafer by using a DI water drop transfer method where the DI water was carefully removed using capillary forces. The procedure was completed by rinsing the wafer with the SWNT film with acetone and isopropyl alcohol and drying at 150 °C in a vacuum of $10^{-6}$ Torr for two hours. A scanning electron microscopy (SEM) image of a typical carbon nanotube network is shown in Fig. 1(a). There is no preferential orientation of CNTs because the vacuum filtration procedure of the film formation leads to random placement of SWNTs.[17, 18] The typical resistivity is ~ 0.002 $\Omega$·cm at room temperature, but it increases by 10-20 times from 300 to 4.2 K.

Figure 1(b) shows a scanning helium ion microscope (HIM) image of a device with a CNT network channel and four magnetic electrodes (F1-F4). A rectangular spin transport channel in Fig. 1(b) is defined by standard e-beam lithography followed by an oxygen plasma treatment. The contact pads consist of Cr/Au (10 nm/90 nm) and the four ferromagnetic leads of $Co_{70}Fe_{30}$ (2 nm)/$Co_{49}Fe_{21}B_{30}$ (10 nm)/ Ta (4 nm)/ TaN (4 nm) on top of the CNT network channel are deposited by magnetron sputtering. As shown in Fig. 1(c) one of the ferromagnetic electrodes, F2 is designed to have a smaller width than the other 3 electrodes (i.e. 200 nm compared to 300 nm) so that it's magnetization switches at a higher magnetic field. Due to the nature of the nanofabrication process, each ferromagnetic electrode may have slightly different switching fields even though they are designed to have the same width. Since the reported spin coherence length of CNTs is ≤ 1 μm, the channel length $L$ in this study is varied from 100 to 800 nm.[1, 4] Among 24 fabricated devices with different channel lengths, 12 devices were tested in the nonlocal geometry and 9 devices showed nonlocal switching hysteresis. Since the nonlocal measurement geometry in Fig. 1(c) completely separates the charge and spin current paths, any spurious magnetoresistance effect is eliminated.[6, 7]

The junction resistance was investigated by using two and four terminal local dc measurements, as shown in Fig. 1(d) and 1(e), respectively. The magnetic field was applied along the easy axis of the ferromagnetic electrodes, which is perpendicular to the current direction in the CNT network channel. At room temperature the junctions show ohmic characteristics, as shown in Fig. 2(a) and 2(c), whereas the current-voltage (*I-V*) curves show nonlinear properties at 6 K, as illustrated in Fig. 2(b) and 2(c). As fitted in Fig. 2(b) the current is found to vary according to a power law of the form $I(V) \propto V^{\alpha}$, with $\alpha \sim 2.56$ (eV < 10 mV) and $\alpha \sim 1.30$ (eV > 10 mV) at 6 K, indicating the tunneling behavior.[19, 20] Strong repulsive interactions of carriers and the Coulomb blockade effect are responsible for the nonlinear current-voltage characteristics, which have been reported in CNT devices at low temperatures.[3, 4, 19] The channel resistance does not scale well with the channel length, as shown in Fig. 2(d), because the contact resistance dominates the total resistance, as can be estimated from the difference between the two and four terminal resistance values.[5, 21]

The pure spin current signal can be probed within the nonlocal measurement geometry which excludes charge current contributions such as those from anisotropic magnetoresistance, the magneto-Coulomb effect, and the anomalous Hall effect in the ferromagnetic leads. If there is the magneto-Coulomb effect, the junction conductance oscillates or linearly scales with the magnetic field.[22] However, our data do not show such a behavior as shown later. The nonlocal resistance using standard lock-in techniques measured at zero bias is defined as R=V/I as shown in Fig. 1(c). The injection current is ~0.5 μA rms at a frequency of 41.7 Hz and a constant baseline resistance is subtracted to get the spin nonlocal resistance ($R_S$). The measurements were performed at 0.25 K in order to overcome the conductance mismatch between the channel and the spin injectors by the natural tunnel barrier formed at low temperatures. As can be seen in Fig. 3(a-e) there is a clear hysteresis effect associated with different magnetic configurations due to the switching of four ferromagnetic electrodes. The different magnetization alignments of the CoFe electrodes are indicated by the vertical arrows in Fig. 3. Starting from 0.5 T, all the ferromagnetic electrodes are aligned along one direction. As the magnetic field is swept from positive to negative values, the lowest resistance state is achieved when F1, F3, and F4 flip their magnetization directions, for example at ~ -0.1 T in Fig. 3(a). In this configuration, the injectors (F1, F2) are antiparallel and the detectors (F3, F4) probe the spatial dependence of the spin down chemical potentials, resulting in a negative nonlocal resistance. As the field is further swept to greater negative values, the magnetizations become parallel again, but pointing in the opposite direction. Ramping up the field to positive values causes the spin signal to increase above the background level at 0.086 T when F1 and F4 reverse. In this magnetization alignment, the detection voltage probes opposite spin directions, with both injectors and detectors being antiparallel, resulting in a positive nonlocal resistance. When F3 flips, the detectors probe a negative nonlocal signal, and finally the magnetization is aligned into one direction when F2 flips.

Some samples show an asymmetric loop with respect to the magnetic field as shown in Fig. 3(b). Such loop shifts towards a negative field can be explained by exchange biasing due to the formation of antiferromagnetic oxides such as CoO and $FeO_x$. Such antiferromagnetic layers below their respective Neel temperatures will give rise to a unidirectional anisotropy of the ferromagnetic electrodes and induces an asymmetric MR signal with magnetic field, as reported previously for multiwalled carbon nanotubes, multilayer graphene, and magnetic tunnel junctions.[3, 23, 24] When there is a strong exchange bias, we only observe a minor loop in which one of the electrodes (F2) is pinned and does not change its magnetization direction, as shown in Fig. 3(c-e).[25]

The staircase like behavior in the nonlocal switching signal of Fig. 3(a) is attributed to the subsequent sequential switching of the ferromagnetic electrodes due to the formation of ripple domains, which typically is found in thin film materials which have a high saturation magnetization such as CoFe used as in our experiments.[8] As the ferromagnetic electrodes consist of many small magnetic grains and different grains are connected to different individual nanotubes, domain switching at different values of the magnetic field contributes to the multi step signal in Fig. 3(a).[3] On the other hand, multiple nanotubes involved in the transport can produce a gradual switching by averaging the various switching fields, as seen in Fig. 3(b). In an ideal case the background nonlocal resistance should be zero because there is no charge current flow. However, there are two types of groups depending on the value of the baseline resistance; one with a large baseline resistance of several $k\Omega$ in Fig. 3(a-c), and the other with a relatively small baseline resistance of ~ 300 $\Omega$ in Fig. 3(d) and 3(e). The finite baseline resistance can be attributed to the finite input impedance if the measurement unit works as a current sink and ballistic carrier motion in the channel.[6, 26] Another report has pointed out that interface inhomogeneity can result in a finite baseline resistance, which is very plausible in our SWNT films.[27] The interplay of Peltier and Seebeck effects to the nonlocal baseline resistance can be ruled out in our samples due to a high resistive channel and a low excitation current.[28]

For tunneling contacts using the general theory of nonlocal measurements, the difference in the magnitude of the nonlocal spin signals for the parallel and antiparallel states of the central two electrodes is given by $\Delta R_s = (P^2 \rho \lambda / A) e^{-L/\lambda}$ labeled in Fig. 3(a), where $P$ is the injected spin polarization, $\rho$ is the resistivity of the CNT channel, $\lambda$ is the spin diffusion length, $A$ is the cross-sectional area of the channel, and $L$ is the channel length.[29-31] The dependence of the nonlocal MR ($\Delta R_s/R$) versus $L$ is plotted in Fig. 3(f) on a log scale, where $R$ is $\rho L/A$. The solid lines are fitted using the above equation, and we obtain values of $\lambda = 1.61$ μm with $P = 0.18$ for the large baseline group (open circles), and $\lambda = 2.45$ μm with $P = 0.41$ for the small baseline group (solid squares). Similar values of the spin diffusion length of 1.4 μm and the spin polarization of 0.2-0.25 have been reported in experiments using a single CNT.[4, 6, 7] There is a reduction in the injected spin polarization at the interface due to the spin flip scattering compared to the spin polarization of 0.55 expected for CoFe.[29, 32] The spin resistance mismatch between the channel and the spin injector is overcome by the nonlinear tunneling characteristics between CNTs and CoFe at low temperatures, leading to spin valve voltages ($V_s$) ranging 50-70 μV in Fig. 3. An upper bound of the spin relaxation time ($\tau_{sf}$) can be estimated to be 32.4 - 75 ps from the relationship of $\lambda = \sqrt{v \tau_{sf} l}$, where the mean electron velocity in the CNTs ($v$) is $0.8 \times 10^6$ m/s and a CNT mean free path ($l$) is 0.1 μm.[20]

Lastly we discuss the origin of spin relaxation in our measurements. In carbon-based materials, the hyperfine interaction is suppressed due to the absence of nuclear spins in $^{12}$C. Thus, the effective spin-orbit coupling (SOC) for conducting carriers is mainly responsible for the spin relaxation, which is the case for our SWNT film devices. Recently, it was suggested that lattice distortion from $sp^2$ to $sp^3$ induced by adatoms chemisorbed on the graphene surface gives rise to the enhancement of the local SOC to the order of 10 meV.[33] This adatom-induced SOC is considered as an origin of the short spin relaxation time of the order of 0.1 ns in graphene,[34] which could also be a possible origin of the short spin relaxation time in our SWNT films, if it is indeed realizable on the SWNT surface. However, previous theoretical works have revealed that SWNTs are rather inert so that adatoms are apt to physisorb onto the SWNT surface than to chemisorb.[35, 36] Hence, the adatoms on the SWNT surface are least likely to induce the SOC. Unlike graphenes, cylindrical curvature of SWNTs affects energy dispersion significantly. Combined with the atomic SOC, the curvature induces the effective SOC of the order of 0.1 meV that was measured experimentally in an ultra-clean SWNT quantum dot.[37, 38] The strength of the curvature-induced SOC depends on the chirality of SWNTs, and the direction of its spin-orbit field is along the SWNT axis.[39, 40] As a result, both strength and direction of the curvature-induced SOC in our SWNT films consisting of randomly oriented CNTs are spatially random with nanoscale variation [see Fig. 1(a)]. In addition, its maximum strength can be enhanced to nearly 1 meV for our CNTs with a diameter of 1.55 nm. Therefore, we believe that the curvature-induced SOC is a primary source causing spin relaxation in SWNT films.

Spin relaxation in SWNT films was not addressed theoretically yet, but several spin relaxation mechanisms arising from the curvature-induced SOC in a SWNT [41-43] and a corrugated graphene [44] were proposed, which could be useful to analyze our measurements. For itinerant electrons in a SWNT, the curvature-induced SOC can generate fluctuating spin precession yielding spin relaxation at room temperature that suppresses as temperature decreases.[43] Also, deformation potential[41] and bending-mode phonons[42] in SWNT quantum dots can cause spin relaxation. The estimated spin relaxation times are at least of the order of 1 μs at 0.1 K for the SOC of the order of 0.1 meV, which could become at least 10 ns for the SOC of order of 1 meV since spin relaxation rate is proportional to the square of the SOC strength. In a corrugated graphene,[44] the curvature-induced SOC of the order of 0.01 meV itself can cause spin relaxation, and its estimated spin relaxation time is about 100 ns. In this mechanism, the spin relaxation time could be about 10 ps for the SOC of the order of 1 meV, which might imply the importance of the random SOC for the spin relaxation in our measurement. In order to estimate a reliable spin relaxation time in SWNT films, it would be required to consider not only the random curvature-induced SOC but also the inter-tube transmission of charge carriers in the presence of the SOC.

In conclusion, we have succeeded in detecting spin transport in a SWNT film over a distance of 1 μm between CoFe contacts in a CNT network using a nonlocal probe configuration to eliminate spurious magnetoresistance effects. The measured nonlocal signal ($\Delta R_s$) from a CNT film is comparable to the largest value of 130 Ω observed in graphene.[45] The estimated spin diffusion length in this film at 0.25 K is 1.61 - 2.45 μm and the injected spin polarization from CoFe into the nanotube network is found to be 18 - 41%. Our observations open the possibility to the fabrication of many carbon nanotube based spintronic devices in a single wafer using a conventional top-down approach by taking advantage of the long spin diffusion length and the high spin injection efficiency without the need for any artificial tunnel barrier in carbon nanotubes. Further experimental works such as the demonstration of the Hanle effect, and room temperature operation using a tunnel barrier and an aligned nanotube film will make the proposed nanotube networks a promising building block for spin transport devices.

This work is partially supported by the Singapore NRF under CRP Award No. NRF-CRP 4-2008-06, the Singapore MOE ARF Tier 2 (MOE2008-T2-1-105), and R-263-000-504-133.

[*] Electronic address: eleyang@nus.edu.sg

**Figure Captions**

Figure 1.  (a) SEM image of a SWNT network. The scale bar is 500 nm. (b) The scanning helium ion microscope (HIM) image of the fabricated device with four ferromagnetic electrodes (F1-F4) and Cr/Au contact pads. (c) The nonlocal spin valve measurement geometry. (d) Geometry of a two terminal local measurement. (e) The four terminal local geometry.

Figure 2.  (a) Current-voltage ($I$-$V$) characteristics of devices with different channel lengths ($L$) at room temperature using a four probe technique. (b) Comparison of $I$-$V$ data using two and four probe techniques from a device with $L$= 200 nm at 6 K. The solid lines in (b) are fits. (c) Resistance versus bias voltage plot. (d) Resistance versus $L$ at various temperatures.

Figure 3.  The nonlocal spin signals with various injector and detector spacing $L$ measured at 0.25 K. A constant baseline resistance of ~5-9 k$\Omega$ and ~300 $\Omega$ is subtracted in (a-c) and (d-e), respectively. The spacing is between the central injector and detector electrodes. (f) The dependence of nonlocal MR on the spacing. The solid lines are fits.

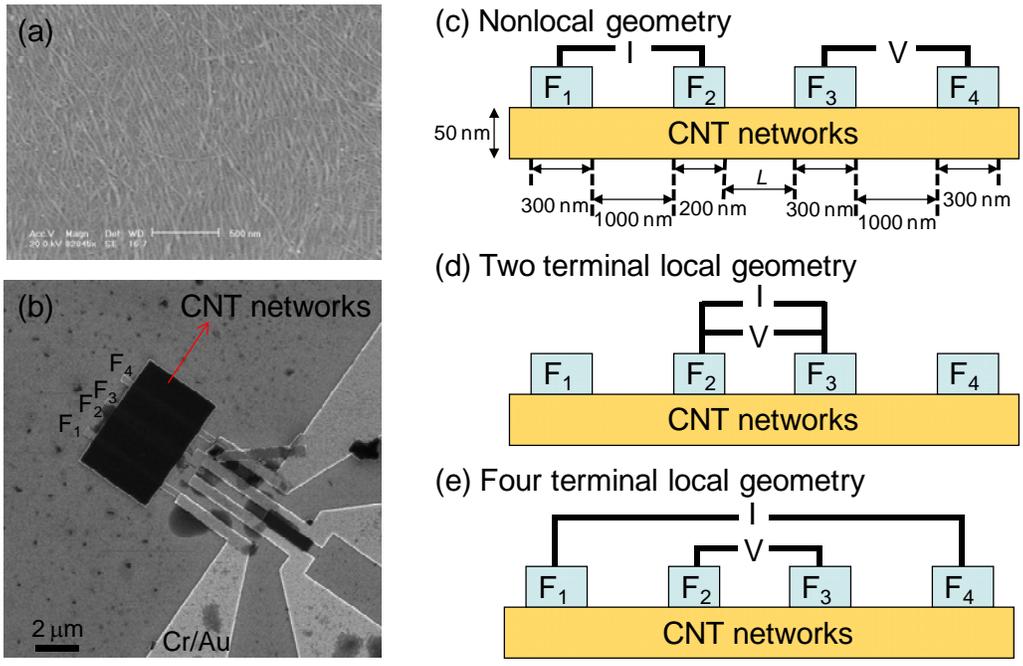

Figure 1

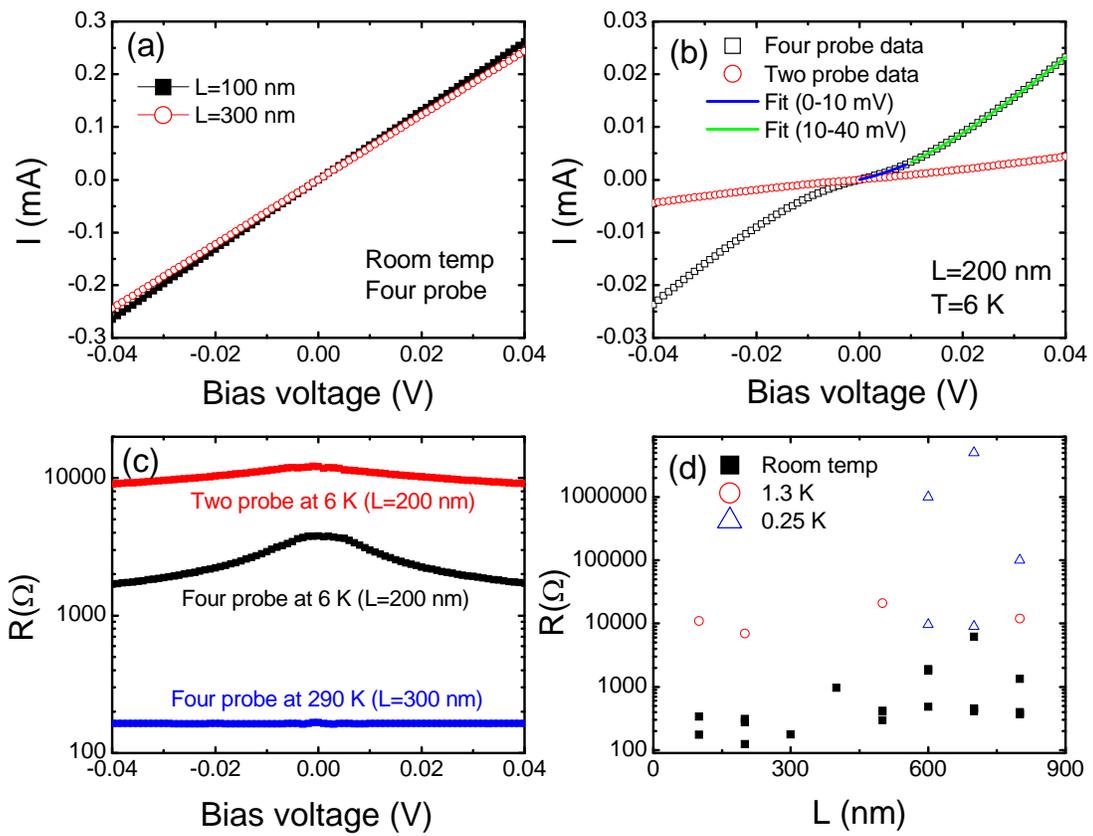

Figure 2

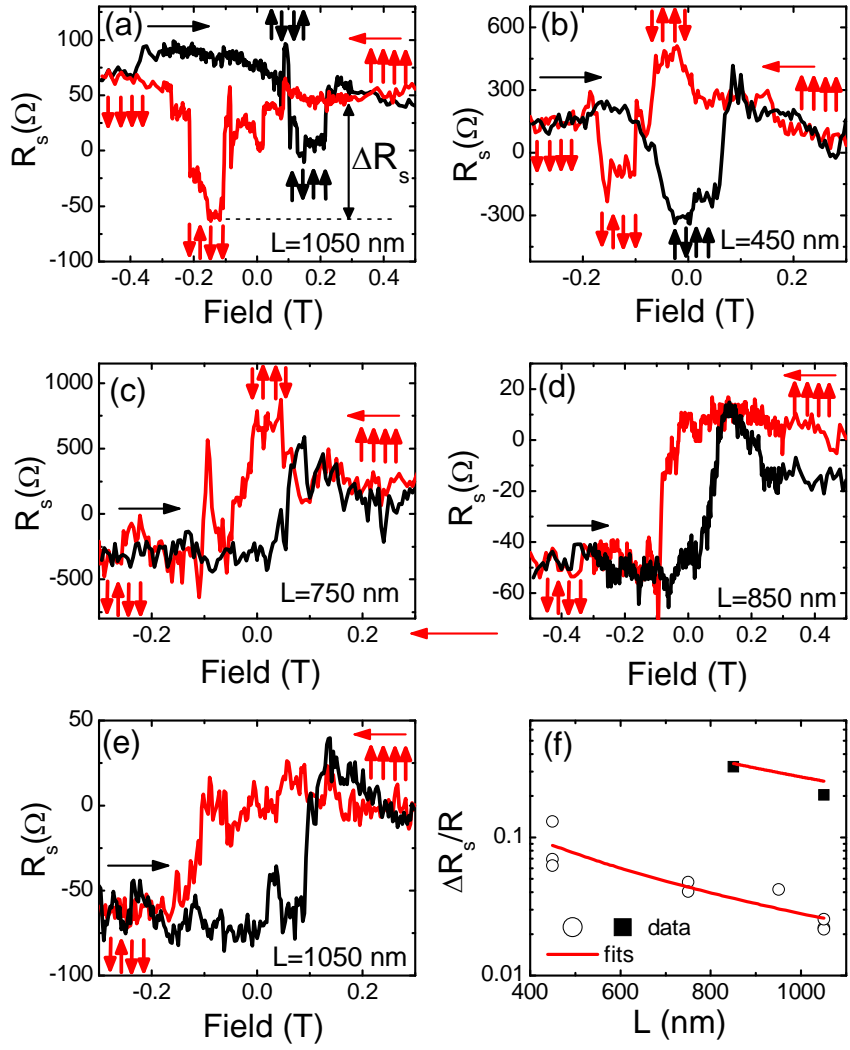

Figure 3